# The Interfacial-Organized Monolayer Water Hindering the Aggregation of Nanographene: Both in Stacking and Sliding Assembly Pathways

*Wenping Lv, Ren'an Wu*[*]

CAS Key Laboratory of Separation Sciences for Analytical Chemistry, National Chromatographic R&A Center, Dalian Institute of Chemical Physics, Chinese Academy of Sciences (CAS), Dalian, 116023, China

**Corresponding Author Footnotes**

Prof. Dr. Ren'an Wu

Tel: +086-411-84379828

Fax: +86-411-84379617

E-mail: **wenping@dicp.ac.cn, wurenan@dicp.ac.cn**

**Abstract:**

A computational investigation was carried out to understand the aggregation of nanoscale graphene with two typical assembly pathways of stacking assembly and sliding assembly in water. The interfacial-organized monolayer water film (MWF) hindering the aggregation of nanographene in both stacking and sliding assembly pathways was reported for the first time. By means of potential mean forces (PMFs) calculation, no energy barrier was observed during the sliding assembly of two graphene nanosheets, while the PMF profiles could be impacted by the contact forms of nanographene and the MWF within the interplate of two graphene nanosheets. To explore the potential physical basis of the "hindering role" of self-organized interfacial water, the dynamical and structural properties as well as the status of hydrogen bonds (H-bonds) for interfacial water were investigated. We found that the compact, ordered structure and abundant H-bonds of the MWF could be taken as the fundamental aspects of the "hindering role" of interfacial water for the hydrophobic assembly of nanographene. These findings are displaying a potential to further understand the hydrophobic assembly which mostly dominate the behaviors of nanomaterials, proteins etc. in aqueous solutions.

**Keywords:** Nanographene, Interfacial-Organized Monolayer Water, Hydrophobic Assembly, Molecular Dynamics Simulation

## Introduction:

It is well believed that the aggregation and subsequent removal of hydrophobic surface from water is to play a critical role in molecular assembly for solutes in water[1] and the hydrophobic hydration dominated the behaviors of hydrophobic molecules in aqueous solutions.[2] While, the hydration thermodynamics of solutes were mostly dependent on their surface topography and size,[1, 3-5] such as the extended surface could modulate the hydrophobic interaction of neighboring solutes[6] and the difference in the physics of hydration of small and large solutes would arise from the different manner in which they affect the structure of water.[5, 7-9] A better understanding of the nanoscale hydrophobic interaction of planar shaped nanomaterials and the corresponding aggregation behaviors in solutions such as particularly in aqueous solutions would not only broaden their applications, but also might therefore yield new insights for potentials in biological systems.

Nanographene or graphene nanosheet, possessing ultrathin thickness, huge surface area, and unique electrical and mechanical properties, is an emerging class of nanomaterials to be applied in various fields (e.g. biosensors, drug carriers and conductive ink etc.).[10-17] The planar shaped graphene nanosheets offer extended interfaces to solvents. The behaviors of interfacial water at the surface of graphene sheets are attracting a great interest for further understanding the hydrophobic hydration as well as the nano-confinement of water molecules.[6, 18-24] Early researches have pointed out the occurrence of density oscillations and molecular orientational biasing of water molecules near the planar hydrophobic surface.[3] For water near carbon-based structure, a thin interfacial layer about 0.5 nm thick could be formed at the water-graphene interface.[21, 25-27] On the other hand, theoretical investigations have also found that the average number of hydrogen bonds per water was reduced at graphene surface, due to the dangling O-H molecular bonds in interfacial water directly pointing to graphene surface.[19, 21, 25] Recently, a direct experimental evidence of the existence of non-H-(non-D) bonded water at water-graphene interface has been observed by means of vibrational spectroscopy.[19] As we know, understanding physical behavior of the interfacial water has important implications for the design of nanostructured devices, as well as the nano-bio interactions.[23, 28-30] Up to date, how the interfacial water impacts the self-assembly of nanographene in water has not been clarified yet.

For hydrophobic particles, the significant fluctuations of the liquid-vapor-like interface surrounding the solutes leading to the formation of a vapor tunnel to accelerate the assembly.[31] Because of the $\pi - \pi$ stacking interaction, the strong attraction between graphene sheets was existed, which leading the direct aggregation of two graphene nanosheets to a stacking form. However, by calculating the potential of mean forces (PMFs) between two graphene sheets, the free energy barriers could be observed in the reducing of the interplate separation of two graphene sheets.[32-34] More recently, Berne et al. further reported that two friction profile peaks were at interplate separation of about 0.88 and 0.62 nm corresponding to the same position of PMF barriers, with two "waiting periods" observed in the fixed stacking assembly for two small graphene-like plates with interplate separation of about 0.95 nm and 0.66 nm, respectively.[34] Actually, with the confinement of two hydrophobic plates such as graphene nanosheets, the structural and dynamical properties of water are different from those of bulk liquid.[21, 22, 27] For instance, the density oscillation of confined water would be enhanced with the narrowing of the separation between two graphene surfaces because of the interfering effect;[21] the diffusivity of interfacial water molecules at the interface was slightly slower than those in bulk-like internal areas but markedly smaller within the interplate due to the nanoscale confinement.[27] These dynamical and structural changes of confined water within interplate of graphene nanosheets might also impact the aggregation of graphene nanosheets in water.

So far, the understanding of the self-assembly behavior of graphene nanosheets in water, to the best of our knowledge, is still unsatisfactory with few studies only.[33, 34] For instance, Aluru et al. investigated the co-aggregation behavior of graphene fragments to understand the formation of graphite-like structure;[33] Blankschtein et al. proposed a kinetic theory of colloid aggregation to quantify the lifetime of suspended graphene in polar solvents.[32] And, Berne et al. pointed out that molecular-scale hydrodynamic interaction was essential in describing the kinetics of assembly of graphene-like plates.[34] Nevertheless, most of these publications focused on the stacking assembly, although the fast combination of graphene nanosheets was observed in *N*-methyl pyrrolidone (NMP) in shear direction after the two sheets collide within a small area.[32] The understanding of the kinetics of the sliding assembly of nanographene in shear direction is rarely reported, particularly the fundamental effect of the hydrophobic hydration shell at graphene surface towards the hydrophobic aggregation of nanographene.

In this contribution, a computational study on the hydrophobic assembly of nanographenes was performed based on the all-atom MD simulations, to understand the interplay between interfacial water and the aggregation behavior of nanographene. Two simulation systems for assemblies in stacking (noted as '||') and sliding (noted as '— —') pathways (Figure 1) were established. By means of the steered molecular dynamics (SMD), the analyses of force spectra and PMFs of nanographene disaggregation in sliding pathway (— —) were carried out. Also, the dynamical and structural properties, including the variation of hydrogen bonds (H-bonds) of interfacial water at different assembly statuses for nanographene were investigated. Moreover, the charge decorating to modulate the aggregation of nanographene was further presented.

In accord with earlier expectation, the translocation and rotation were observed during the self-assembly of nanographene in stacking pathway (||). For sliding assembly (— —), the energy barrier was not observed in PMFs, but the contact form of two graphene nanosheets could impact the time of aggregation. Interestingly, an interfacial-organized monolayer water film (MWF) could be spontaneously formed within the interplate of nanographene in stacking and even in sliding assembly pathways. The study of structural properties of MWF shows that the abundant H-bonds were existed within MWF. It suggests that a networked H-bonds on the extended graphene surface was formed, similar to the H-bonds network on small hydrophobic solutes surface.[8] Though the graphene nanosheets finally adhered directly due to the dynamical fluctuation of the MWF and attraction between graphene, the charge decoration at the corners of graphene nanosheets could effectively stabilize the MWF by means of preventing the contact of graphene edges.

**Simulation details and methods:**

The graphene nanosheet used in self-assembly simulations was sized as 4.72 X 4.92 nm$^2$. Each sheet consists of 880 carbon atoms with the dangling bonds at graphene edges capped by 84 hydrogen atoms. To study the stacking assembly of graphene nanosheets, two congruent graphene nanosheets were arranged in stack status (||) initially (Figure 1a). The interplate distance of *d* was 2 nm. The size of the simulation box was 8.0 X 8.0 X 6.0 nm$^3$. As diagrammatized in Figure 1b, two in-planed graphene nanosheets (— —) were used to study the sliding assembly of graphene. The edge distance of $d_i$ between two graphene nanosheets was considered as an impact factor to the sliding assembly of nanographene. Four simulation systems with different $d_i$ (0.3, 0.6, 0.9 and 1.2 nm) were set up. The dimensional sizes of four simulation boxes were 12.0 X 10.0 X 4.0 nm$^3$, 12.0 X 10.0 X 4.0 nm$^3$, 12.3 X 10.0 X 4.0 nm$^3$, 12.6 X 10.0 X 4.0 nm$^3$, respectively. The parameters for other MD simulations were declared in discussion section. All the simulation boxes were fully filled with TIP3P water molecules.

The potential of mean forces (PMFs) for sliding assembly of graphene nanosheets were constructed from SMD simulations based on Jarzynski's equality. [35, 36] The pulling velocity of 0.005 nm/ps applied in this work was over 20 times slower than the velocity of sliding assembly of graphene nanosheets. The external work was calculated by integrating the force over the pulled distance from SMD trajectory:

$$W(r) = \sum_{k=0}^{n} \int_0^r F_k(r')dr' \quad (1)$$

Where *n* is the number of pulling groups, $F_k$ is the pulling force applied on the *k*th pulling group.

The second-order cumulate expansion of Jarzynski's equality was used to derive the PMF or free-energy difference from the work *W* as follows:

$$\Delta G = <W> - \beta \sigma_w^2/2 \quad (2)$$

Where $\beta = 1/k_b T$, T is the temperature and $k_b$ is the Boltzmann constant. <*W*> is the mean work averaged from all trajectories and $\sigma_w$ is the standard deviation of the work distribution. 11 SMD simulations from different initial conformations were performed for each PMF calculation.

All MD simulations were performed using GROMACS 4.5 program package. The AMBER99 force field was employed to model water molecules. The parameters for graphene carbon atoms were those of $sp^2$ carbon in benzene in AMBER99 force field. The cut-offs of the van der Walls (vdW) force were implemented by a switching function starting at a distance of 1.1 nm and reaching to zero at 1.2 nm. The particle mesh Ewald (PME) method was used to calculate the electrostatic interactions with a cut-off distance of 1.4 nm. Three-dimensional periodic boundary conditions (PBC) were applied in simulation. Time step of 2 fs was set. For each simulation, 1000 steps energy minimization and 500 ps solvent relaxation before the production simulation were performed.

## Results and discussion:

### Stacking self-assembly of graphene nanosheets

Three dimensional distance evolution of two graphene nanosheets and the snapshots of typical conformations are shown in Figure 2. The total assembly time of two graphene nanosheets was about 1.4 ns. During the first 1 ns of assembly, the average interaction energy between graphene nanosheets was -0.145±0.495 kJ/mol (Figure S1a, Supporting Information). It was extremely low as compared to that (over -4650 kJ/mol, Figure S1a) of the stacking interactions between two completely assembled graphene nanosheets, suggesting the domination of molecular diffusion in this process.[32] The attraction-interaction-induced aggregation (effective aggregation) of two graphene nanosheets was achieved less than 400 ps. The pronounced fluctuation of dx and dy curves ($\Delta x$, $\Delta y$) in Figure 2 shows that the complicated orientational variations of nanographene were undergone before the effective aggregation. Two kinds of typical orientational adjustments during the diffusion process were the in-plane rotation (t=560 ps, Figure 2) and the shear shift (t=1040 ps, Figure 2) between two graphene nanosheets. These orientational variations seemed to maximize the degree of disorder between graphene nanosheets but minish the disturbance of water environment, undergoing as a thermal diffusion determined process.[32] However, the interplate separation of nanographene was not reduced obviously in this period (see dz curve and its variation($\Delta z$) in Figure 2). It suggests that the damping of graphene migration in normal direction was greater than that migration in shear direction (in-plan rotation and/or shear shift). On the other hand, the thermal movement of graphene nanosheets also ruined the parallel orientation between graphene nanosheets. As shown in Figure 2, a further reduction of the interplate separation lead to one side of the graphene nanosheet was approaching with another in advance (t=1040 ps, Figure 2). Interestingly, the hydration shell within the region that two hydrophobic graphene nanosheets approached were not expelled out. The existence of these interlamination interfacial water molecules induced the interaction energy between graphene nanosheets was increased mildly (Figure S1a, Supporting Information). After a short duration, a graphene-water-graphene sandwiched structure (GWGSS) was spontaneously formed by free self-assembly at last (t= 1280 ps, Figure 2). The interplate separation between two planes of GWGSS was maintained steadily for about 0.66 nm before the further aggregation (see dz curve in Figure 2), which indicated that the confined interfacial water was composed of the monolayer water molecules.[27, 32, 34] The interaction energy between two graphene nanosheets was stayed at -450 kJ/mol at this stage, which was much weaker than the interaction energy between two directly stacked graphene nanosheets (over -4650 kJ/mol, Figure S1a, Supporting Information). These results suggest that the interaction between graphene nanosheets was effectively decoupled with the existence of the confined monolayer water film (MWF). However, the existence of MWF was maintained only for about 90 ps (marked as orange band in Figure 2) in this self-assembly process. Once a part of the graphene nanosheets (t=1384 ps, Figure 2) adhered together directly, the MWF was extruded out very quickly (less than 20 ps). This observation suggests that the stacking assembly of nanographene hindered by the interlamination interfacial water with a "two-step" assembly process might exist in the hydrophobic assembly of nanographene (Movie S1, Supporting Information).

By means of freezing the freedom degree of one graphene nanosheet, the self-assembly process was changed as a flexible graphene nanosheet adsorbed onto a flat graphene substrate in aqueous solution.[34] We found that in such a fixed-stacking assembly process, the survival time of MWF was prolonged to the length of 800 ps. It was almost 10 times of that observed in free self-assembly (90 ps) and near to one third of the total assembly time (about 2.8 ns), showing that the stability of MWF was enhanced in the adsorption of nanographene on a flat surface. It seemed that the flexibility reduction of nanographene could promote the self-organizing of interfacial water. This study was consistent with the observation of the hindering of water molecules but not the monolayer interfacial water towards the fixed-stacking assembly of smaller graphene-like plates (about 1 nm).[34]

**Sliding self-assembly of graphene nanosheets**

In sliding assembly of nanographene, the edge distance between two graphene nanosheets was considered as the impact factor to the self-assembly. Four simulations with different initial edge distances ranged from 1.2 nm to 0.3 nm were carried out. The total time and the effective time of the sliding assembly processes were plotted as a function of the initial edge distance of graphene nanosheets in Figure 3a. Where, the total time was defined as the time period from the beginning to the end of self-assembly in simulation. The starting point of the effective time was defined as the position where the interaction energy between two graphene nanosheets was weaker than -200 kJ/mol with about 5% overlapped area of two graphene nanosheets. As shown in Figure 3a, the graphene nanosheets in those simulations were aggregated in a very short time. Both the total time and the effective time of the sliding assembly were increased with the increase of initial distances between two graphene nanosheets. The increase of total time for the diffusion processes of graphene nanosheets seems to be reasonable and related with their initial distances.[32] The prolonged diffusion process also induced the orientations of nanographene disordered.[32] As shown in the insets of Figure 3a, the contact conformations of graphene nanosheets in those simulations seemed varied with the initial edge distance. Subsequently, the sliding assembly of graphene nanosheets accompanied with in-plane rotation, to maximize their attraction energy.[32] These results suggest that the contact conformations of nanographene and the induced in-plane rotation could impact the effective time of nanographene aggregation (further discussions presented in the section of *Force spectra and PMFs of sliding assembly of nanographene*).

As we know, the separation in normal direction of two graphene could impact the aggregation behavior of nanographene in sliding assembly. Considering a situation that the two graphene nanosheets were arranged side by side but with a separation in normal direction (Figure 3b), the assembly of graphene nanosheets should be still maintained in sliding modality for their small separation in normal direction. A model simulation was carried out to verify this suppose. The separation distance of two graphene nanosheets in *z*-direction was 0.7 nm (slightly bigger than the interplate separation distance of two graphene in GWGSS, 0.66 nm). Though no GWGSS formed in three of the five MD simulations, graphene nanosheets either stacked on each other directly at the beginning of simulation or the MWF collapsed at the half way of assembly, the successfully formed GWGSS could be observed in the other two assembly processes of graphene nanosheets. The critical snapshots for the formation of MWF during a sliding assembly of graphene nanosheets were shown in Figure 3c-3d (Movie S2, Supporting Information). It indicated that the interfacial-organized water molecules on graphene surface could sustain the attraction interaction between graphene nanosheets even accompanied with the relative-sliding of graphene. The effective assembly time of graphene nanosheets were prolonged to 273.4 ps and 374.4 ps for these two repeated simulations, respectively (Figure S1b, Supporting Information), which were many times of that for direct-contacted sliding assembly (less than 50 ps). This might be attributed to the interaction between graphene nanosheets decoupled by the existence of the MWF (Figure S1b, Supporting Information). Similar to the stacking assembly, the GWGSS was not stable and would be broken down after a short duration. These findings suggest that the "two-step" assembly also occurred in the sliding assembly pathway.

**Force spectra and PMFs of sliding assembly of nanographene**

To explore the fundamental mechanism of the sliding assembly, the force spectrum analysis and PMFs calculation were further performed. As we pointed previously, the contact conformation and the separation of graphene in normal direction ($d_z$) might impact the aggregation of two graphene nanosheets. Here, the contact angle ($\theta$) between two graphene nanosheets was used to measure the relative orientation of graphene nanosheets. The force spectra and PMFs for three typical assemblies including the edge-edge contacted assembly ($\theta = 0º, dz = 0$ nm), edge-corner contacted assembly ($\theta = 45º, dz = 0$ nm) and indirect-contact assembly ($\theta = 0º, dz = 0.7$ nm) were studied. The SMD simulation was used to investigate the disaggregation (the inverse process of aggregation) of nanographene.

The pulling groups and the direction of pulling forces were diagrammatized as the insets in Figure 4a-4c.

The time evolution of pulling forces for the two direct-contact assembly pathways was shown in Figure 4a (edge-edge assembly) and 4b (edge-corner assembly), respectively. It shows that the pulling forces acted on each pulling group were synchronously varied within the disaggregation processes in edge-edge assembly pathway (Figure 5a). However, a significant difference occurred (around 600 ps, Figure 5b) among the pulling forces acted on each pulling groups in edge-corner assembly pathway, indicating that the edge-edge contacted sliding assembly of two graphene nanosheets in water was so stable that no observable in-plane rotation happened between graphene nanosheets. While, the asynchronous variation of the pulling forces on each pulling groups in edge-corner pathway indicates that the sliding assembly of two edge-corner contacted graphene nanosheets may be along with the in-plane rotation.

The PMFs and the intermediate states of graphene nanosheets in two contacted assembly pathways were shown in Figure 4d-4e. For edge-edge contacted assembly pathway, three PMFs were calculated at different simulation temperatures (T=285 K, 300 K and 315 K). As plotted in Figure 4d, three PMFs were almost overlapped. The enlarged view of PMFs was illustrated as the inset in Figure 4d. It indicates that the free energy fall was slightly increased with the rise of simulation temperature. However, it is negligible as compared to the total free energy fall of graphene nanosheet aggregation in edge-edge pathway (about -4600 kJ/mol). Different from the stacking assembly,[33] these results suggest that the entropic contribution to the free energy change in sliding assembly pathway was negligible. In other words, the disturbance of water environment in sliding assembly was significantly reduced as compared to that in stacking assembly. Besides, the slopes of PMFs were almost constant until these two graphene nanosheets were completely separated (over 5 nm in reaction coordinate). It indicates that there was no energy barrier existed during the sliding assembly of nanographene. As shown in Figure 4e, the slope of PMF in edge-corner pathway was slowly decreased with the increase of reaction coordinate, suggesting that the interaction between nanographene was reduced with the increase of separation. Additionally, the free energy drop was about -3857 kJ/mol in the edge-corner assembly pathway. It was about 800 kJ/mol lower than that of in the edge-edge assembly pathway. In the inset of Figure 4e, the average interaction energies between two graphene nanosheets were plotted as a function of contact angle. The attraction of two graphene nanosheets was reduced unless the two graphene orientated in the edge-edge contact forms (contact angle was 0°, 90° and 180°, respectively). Thus, the edge-corner contacted graphene nanosheets were trend to rotate for achieving the stable edge-edge contact form with lower free energy during the aggregation. These results show that the contact angle ($\theta$) between two graphene nanosheets impacted the free energy profile of nanographene aggregation.

Both the pull forces and free energy drop of the indirect-contact ($d_z$=0.7 nm) assembly (Figure 4c, 4f) of nanographene were significantly numerical lower than that of the direct-contacted assembly. Moreover, the pull forces applied on different pull groups were fluctuated obviously (Figure 4c). It indicates that the probability of the in-plane rotation of nanographene during indirect-contact sliding assembly was magnified (compared to the direct-contacted assembly). The free energy drop during the sliding assembly process was only about -800 kJ/mol (Figure 4f). It is acceptable because the attraction between two graphene nanosheets was significantly decoupled (about -450 kJ/mol, Figure S1b, Supporting Information). Excluding the interaction energy between graphene nanosheets, the rest of the free energy drop could be considered as the free energy change of water. These results suggest that the existence of MWF within the interplate of nanographene significantly impact the free energy profile during the aggregation of nanographene in the indirect-contact sliding assembly pathway.

**The role of interfacial water to the assembly of nanographene**

The observed "two-step" assembly process of graphene nanosheets in both stacking and sliding pathways shows that the interfacial water played an intricacy role in the hydrophobic assembly of nanographene. It seems inconceivable that the water molecules could be spontaneously arranged within the interplate of hydrophobic surfaces, especially in the sliding assembly pathway. Therefore, the evolutions of hydration shell during the sliding assembly of nanographene, along with the dynamical, structural and H-bonds properties of the MWF within interplate of nanographene were investigated to explore the role of interfacial water to nanographene aggregation.

The hydration shell of graphene in water could be considered as the first layer of the compact interfacial water nearby graphene surface.[25] The hydration shell of graphene edge was much similar to that on the surface of small hydrophobic molecules (see Figure 5a).[8] In this work, three MD simulations were carried out to probe the role of interfacial water on nanographene aggregation in sliding pathway. Due to the flexibility of nanographene during aggregation, one of the graphene nanosheets was restrained at its initial position to improve the observability of the interfacial water in MD simulations. Another free motional graphene nanosheet was initially positioned at the location where the edge-edge distance in *x*-direction ($d_x$) was 0.3 nm, but in z-direction ($d_z$) was 0.0 nm, 0.4 nm and 0.7 nm, respectively. As shown in Figure S2 (Supporting Information), for these three initialized orientations corresponding to the situations of no hydration shell, only one hydration shell and two hydration shells were formed between nanographene, respectively.[25]

When the separation of two graphene edges was 0.3 nm, no MWF was formed. The critical conformations of hydration shell during the assembly were plotted in Figure S3a (Supporting Information). It shows that the vacuum-like interface between the edges of two graphene nanosheets made the nanographene easy to contact. Due to the huge stacking interaction between graphene nanosheets, two contacted graphene nanosheets could be directly aggregated. As a comparison, the MWF could be formed for $d_z$=0.7 nm. The snapshots of the formation of MWF during the sliding assembly of graphene nanosheets show that two graphene nanosheets could smoothly slid on the interfacial-organized monolayer water molecules (Figure S3b, Supporting Information). These results also conform to that of free-assembly in sliding assembly pathways (Figure 3).

The most interesting results were of the separation of two graphene nanosheets at 0.5 nm. In Figure 5b, the conformational variation of hydration shell during the "contact" process and subsequently sliding assembly were displayed. It could directly show that the hydration shells at the contact region (highlighted in yellow) would hinder the aggregation of graphene nanosheets. Specifically, although two graphene nanosheets "attempted" to "touch" each other several times, the presence of "H-bonds-networked water" (Figure 5a) around the graphene edges hindered them. After several rounds of "contacting" between graphene nanosheets, a small overlapped region was formed at the margin of graphene (T=1160ps, Figure 5b). After that, two graphene nanosheets were collapsed to share a common monolayer interfacial water (T=1300ps, Figure 5b) due to the stacking interaction. Similar with other indirect contacting sliding assembly, the GWGSS was subsequently formed. The dynamic evolution of interfacial water during the sliding assembly of nanographene was shown as video in Movie S3 (Supporting Information). These results suggest that not only the confinement but also the interactions within the interfacial water (perhaps H-bonds) played as a critical role in the formation of MWF.

Due to the short life time of GWGSS in free assembly processes, two restraint MD simulations were carried out to capture the dynamical, structural and H-bonds properties of MWF: 1) 1D restraint MD (the motion of graphene nanosheets in z-direction was restrained by a harmonic potential); 2) 3D restraint MD (the motion of graphene nanosheets in *x-*, *y-* and *z*-directions were restrained by a 3D harmonic potential). The force constants of those harmonic potentials were set to 1,000 kJ/(mol · nm$^2$).[37, 38] For 1D restraint MD, only relative sliding in *x*- and *y*-direction could happen between graphene nanosheets. Thus the collapsing of MWF could be prevented during simulation (20 ns). While in 3D restraint MD, two graphene nanosheets were fixed in their initial positions and no relative sliding occurred. The time evolution of the amount of retentive water molecules was used to characterize the dynamics of MWFs. The retentive water molecules were those which stayed in the confined region for the entire interval of time between t and t+*Δ* t. Where,*Δ* t was chosen to 2.5 ns.

Figure 6 shows the evolution of the number of retentive water molecule with time for MWF and the statistical distribution of the retention time. Different with the solid-like water monolayer under graphene cover on hydrophilic substrate,[20, 24] results show that most of the water molecules (over 90%) in MWF were dissipated within the first 450 ps in both 1D and 3D restraint simulations (in Figure 6a and 6b, respectively). Whereas, completely supplanting the rest of water molecules within the confined region by the water molecules in environment would take a longer time (over 2 ns). It suggests that the diffusivity of MWF within the interplate of graphene nanosheets was still maintained. In other words, the forming and breaking of H-bonds frequently happened within MWF.[39] It might be an important incentive of the instability of MWF in free MD simulation. The statistical distributions of retention time show that the most probable retention times were about 900 ps and 1200 ps for 1D and 3D restraint simulations, respectively (see insets in Figure 6). It seems that the water molecules within a fixed slit (3D restraint) were less diffusible as compared to the slidable slit (1D restraint). However, the range of retention time of water in slidable slit (530 ps to 2086 ps, inset of Figure 6a) was significantly broader than that in fixed slit (774 ps to 1848 ps, inset of Figure 6b). It caused by the "tailing effect" of the retention time was significant in the sliding slit case (Figure 6a). These results indicate that the relative sliding of two graphene nanosheets could impact the dynamics of MWF.

The molecular orientation and density were employed to characterize the structural properties of MWF (Figure 7a). Here, the molecular orientation of water molecules, $\Phi$ was defined as the angle between the $H_1$-O-$H_2$ plan and the graphene nanosheet (see the inset of Figure 7b). The normalized orientation distribution of MWF (accumulated from the trajectory of 1D restraint MD simulation) was plotted with triangle in Figure 7b. It shows that the most probable orientation of water was about 13°, and over 60% water molecules were orientated within 30° (marked with red lines in Figure 7b). It indicates that most of water molecules within MWF were orientated in-plane at room temperature. However, the molecular orientation distribution of unconfined interfacial water (plotted with blue dots in Figure 7b) showed that there was no peak within 30°. The interfacial water could not spontaneously be orientated in-plane. This indicates that the structure of hydration shell was transformed due to the confinement of nanographene. On the other hand, the density oscillation of water was also observed in the GWGSS (Figure 7c). It is similar to the density profile of water nearby hydrophobic plane.[25, 40] The density of interfacial water was about 1.5 g/cm$^3$. The MWF owns the highest density, over two times of that in block water. The high density of these hydration shells could be another reason for that interfacial water hindering the aggregation of nanographene.

Furthermore, the status of H-bonds of MWF and the first layer of interfacial water at the outer surface of graphene (unconfined interfacial water), along with a block water slice (Block water, dimension was 4.0×4.0×0.35 nm$^3$) were analyzed. Hydrogen bonds were determined based on cutoffs for the O···H—O angle larger than 145° and the O···H distance shorter than 0.35 nm. The status of H-bonds was described by the internal H-bonds (the number of H-bonds within the analyzed water layer), external H-bonds (the number of H-bonds between the analyzed water layer and the surrounded water molecules) and average number of H-bonds per water ($<n>$).

The average number of H-bonds per water ($<n>$) was widely used to describe H-bonds status of water.[3, 21, 27] Similar to previous studies, the $<n>$ of block water was around 3.28, and decreased at the interface of extended hydrophobic surface (3.01±0.08).[21] However, result shows that the $<n>$ of MWF (2.83±0.08) was also lower than that of block water (3.28±0.07). It seems disaccord to our discussion in previous. While, we also noticed that the number of H-bond within MWF (internal H-bonds) was 243.86±10.06. It's even more abundance than that within a block water slice (233.10±12.62), which contained more water molecules. The internal H-bond of interfacial water on the outer surface of graphene was fluctuated around 207.33. It was obviously decreased as compared to that in MWF. These results confirm that the H-bonds were abundant within MWF and an integrated 2D H-bonds network could be formed (Figure 7d-7e). Therefore, the compact, ordered structure and abundant H-bonds of MWF could be taken as the fundamental reasons of the

"two-step" assembly of nanographene.

**The stabilization of the graphene-water-graphene sandwiched structure**

Tracing the MD trajectories of nanographene aggregation, it shows that the collapse of GWGSS usually began from a bended-adherence of the corners of graphene nanosheets. And the approached region of two graphene nanosheets rapidly extended due to the huge attraction between them. During this process, the water molecules were extruded from the slit and only the H-bonds at the adhering interface between graphene nanosheets were destroyed (Movie S4, Supporting Information). The process of the breakdown of MWF suggests that averting the contact between graphene edges could prevent the collapse of GWGSS. There were many strategies to avert the contact of graphene boundaries, such as introducing the electrostatic repulsion by charge decoration, or introducing the steric hindrance by chemical modification on the graphene edges. In this work, the enhanced stabilization of MWF by charge decoration of graphene nanosheets was demonstrated.

As the adhering of graphene nanosheets usually induced by the bending of nanographene corners due to their flexibility, carbon atoms at four corners of graphene nanosheets were decorated with a static charge of 0.5$e/C$.[30] MD simulation revealed that the collapse of GWGSS was successfully prevented by the electrostatic repulsion, and the MWF could be steady existed within the interplate of graphene nanosheets during the whole simulation duration (10 ns, Movie S5, Supporting Information). It suggests that decoration of graphene corner and/or edge may be utilized to stabilize the MWF within interplate of two graphene nanosheets.

## Summary and Conclusion:

The dynamical evolutions of stacking assembly for nanographene show that the interfacial water could be briefly maintained within the confinement of two graphene nanosheets. By using a fixed flat graphene nanosheet, the survival time of MWF was obviously prolonged. For sliding assembly of nanographene, the aggregation might be impacted by the contact forms between graphene nanosheets. The MWF could be spontaneously formed in sliding assembly that there was a separation initially existed between two graphene nanosheets in normal direction. These results suggest that the "two-step" aggregation could be presented in either stacking or sliding self-assembly of nanographene.

By means of SMD simulation, the fundamental mechanisms of the aggregation of nanographene in sliding assembly pathway were explored. We found that the force spectra and PMFs of graphene nanosheets disaggregation were related to the contact forms between graphene nanosheets. For the direct-contacted sliding assembly of nanographene, no free energy barrier was observed in PMFs. The temperature-insensitivity of PMFs in the direct-contacted sliding assembly suggests that the disturbance of water environment in sliding assembly was significantly reduced as compared to stacking assembly. For the indirect-contact sliding assembly, the free energy drop was numerically reduced due to the MWF within the interplate of two graphene nanosheets decoupled the interactions between nanographene.

The evolution of hydration shell during sliding assembly directly shows that the interfacial water layer played a "hindering role" in the hydrophobic assembly of nanographene. Moreover, the dynamic property analysis suggests that the diffusivity of water molecules in MWF was still maintained, but impacted by the relative sliding of two graphene nanosheets. The investigation of structural and H-bonds properties of the interfacial water layers within the interplate and on the outer surface of nanographene indicates that an H-bonds network within the highly-ordered MWF was formed. These results suggest that the compact, ordered structure and abundant H-bonds of MWF could be taken as the fundamental reasons of the

"two-step" assembly of nanographene.

Although the interfacial water layer could hinder the aggregation of graphene nanosheets, the instability of H-bonds, the flexibility of graphene and the huge stacking interaction between graphene finally induced the breakdown of GWGSS. However, based on above investigations, we found that the stability of the fragile GWGSS could be enhanced by means of charge decoration on graphene corners.

In summary, the MD simulations at atomic level reveal that the interfacial-organized, highly-ordered monolayer interfacial water on the surface of nanographene would hinder the aggregation of planar shaped nanographene in water. Actually, a similar behavior of water molecules was that water could diffuse and exist within the subnanometer scale hydrophobic cavities of carbon nanotubes (CNTs).[41-46] The 2D interfacial-organized monolayer water within the interplate of two hydrophobic graphene nanosheets could be explained as an expanded property of the unique behavior of water molecules on the $sp^2$-carbon family molecules. Our findings could be useful in the rapidly developed nanotechnologies, as well as the further understanding of the hydrophobic interaction dominated aggregation of bio-systems such as proteins.[8, 20, 47-51] Future studies might focus on exploiting the applications of these unique behaviors in self-assembly of more complicated nanostructures, including the interactions between biomolecules and nanomaterials.[52, 53]

## Acknowledgement:

This work was supported by the National Natural Science Foundation of China (No. 21175134), the Knowledge Innovation Program of Dalian Institute of Chemical Physics and the Hundred Talent Program of the Chinese Academy of Sciences to Dr. R. Wu.

**Supporting Information:**

The evolution of interaction energy for the assembly of two graphene nanosheets in stacking (a) and sliding (b) pathway was plotted in **Figure S1**. The initial orientation of graphene nanosheets in three simulations (edge-edge distance in *x*-direction ($d_x$) was 0.3 nm, but in z-direction ($d_z$) was 0.0 nm, 0.4 nm and 0.7 nm, respectively) were shown in **Figure S2**. The snapshots of the evolution of hydration shells during the sliding assembly of nanographenes were shown in **Figure S3**, with the separation of two graphene nanosheets in z-direction is (a) 0 nm and (b) 0.7 nm, respectively. The process of two graphene nanosheets assembly in stacking pathway was shown in **Movie S1** as video. The process of two graphene nanosheets (with a separation of 0.7 nm in normal direction) assembly in sliding pathway was shown in **Movie S2** as video. The dynamical evolution of interfacial water during the sliding assembly of nanographene was shown in **Movie S3** as video. The process of extruding the monolayer water film (MWF) out of the interplate of two graphene nanosheets was shown in **Movie S4** as video. **Movie S5** displays that the graphene–water-graphene sandwiched structure was successfully maintained during a 10 ns MD simulation.

**Figures and Tables**

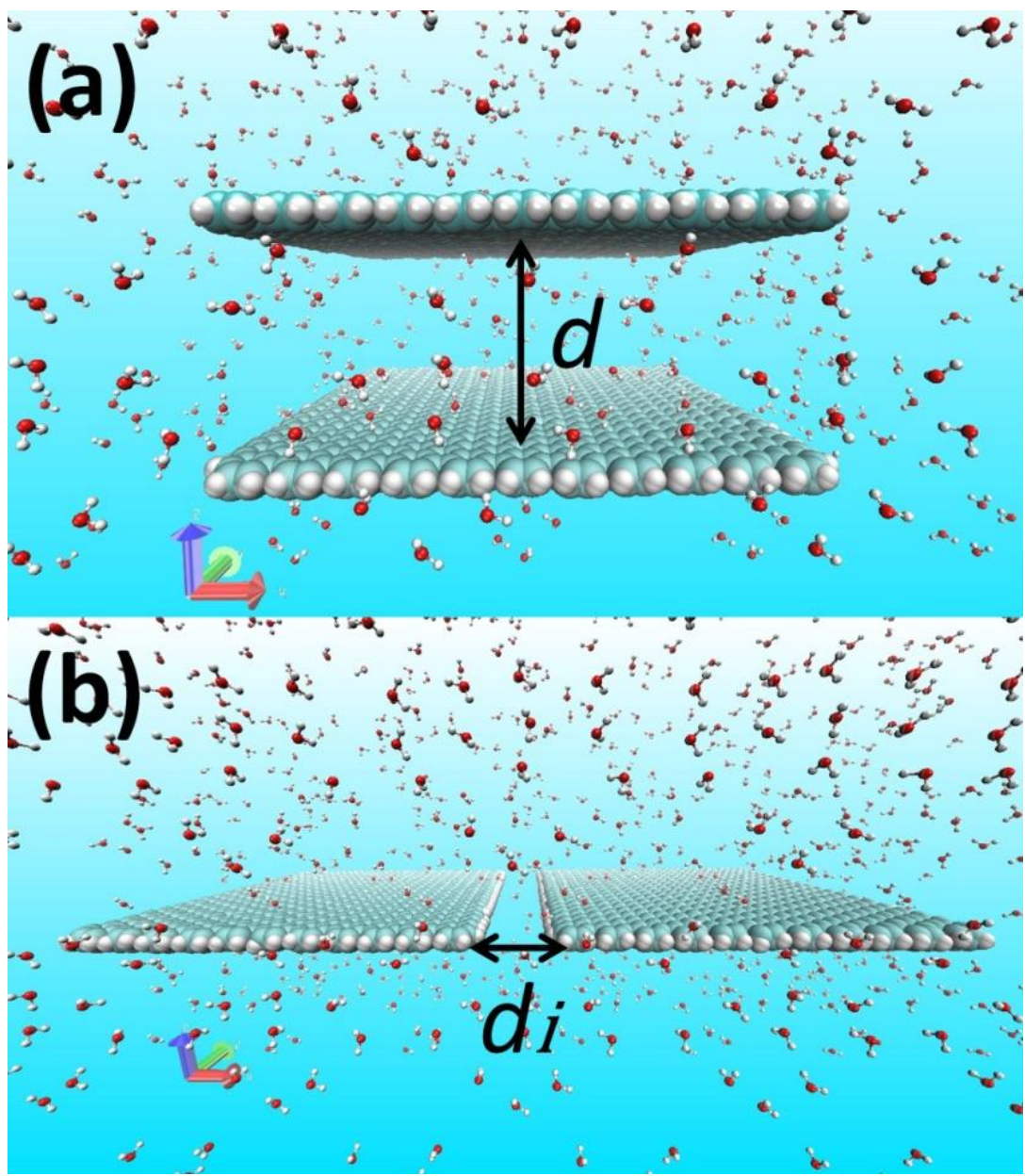

Figure 1. Molecular visualization of two graphene nanosheets for self-assembly in water with initial (a) stack and (b) parallel status. Symbols of $d$ and $d_i$ represent (a) the interplate separation of two graphene nanosheets and (b) the edge distance of graphene nanosheets (b), respectively. In this work, $d$ is equal to 2 nm; $d_i$ is in the range of 0.3 to 1.2 nm.

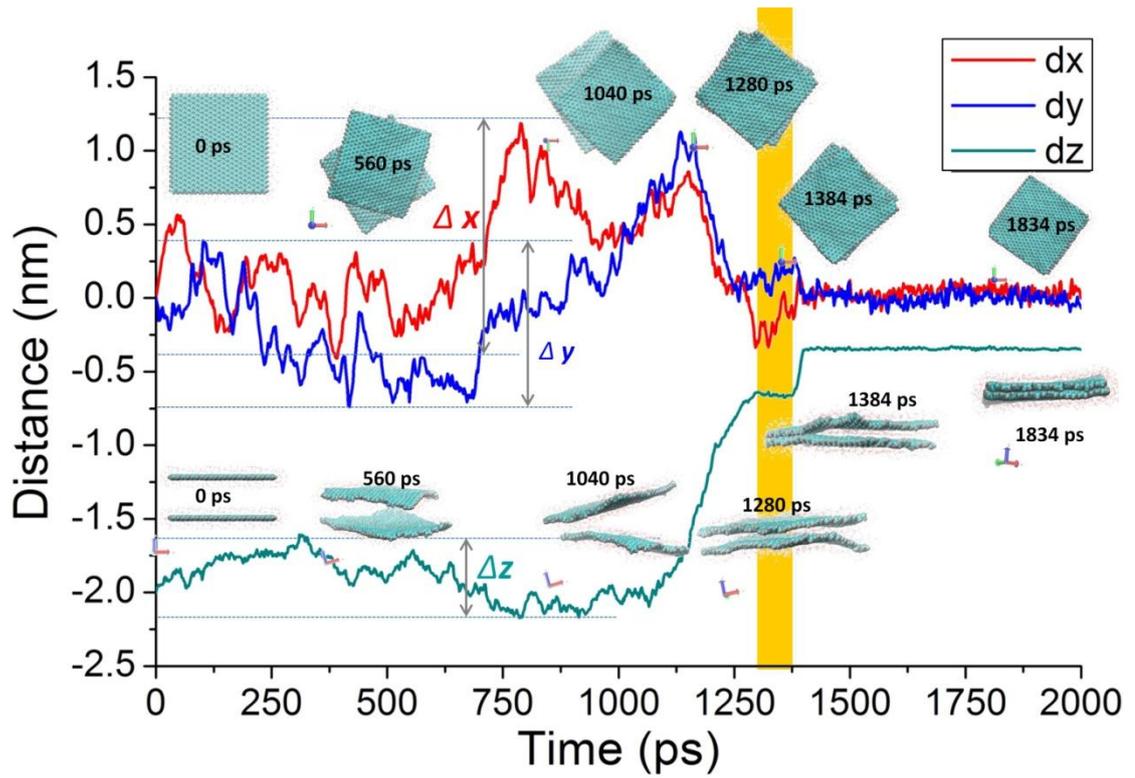

Figure 2. The evolution of three dimension distance for two graphene nanosheets with conformation snapshots during stacking assembly. The distance fluctuations in shear and normal directions were marked as Δ x, Δ y and Δ z, respectively. The width of orange band indicates the duration of graphene-water-graphene sandwiched structure (GWGSS).

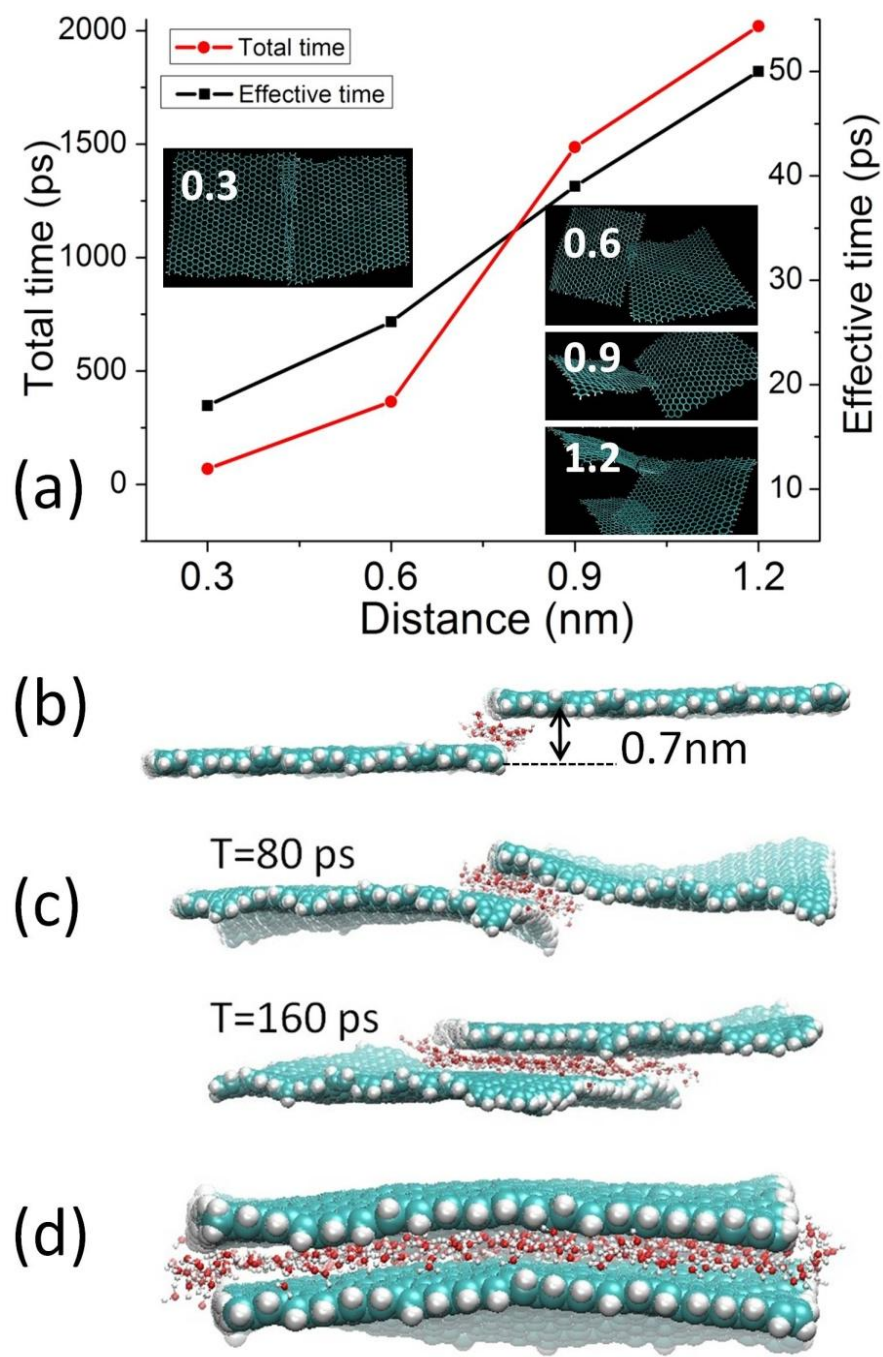

Figure 3. (a) The time evolution for the aggregation of two graphene nanosheets in sliding assembly with different edge distances. The contact conformations of two graphene nanosheets were shown as insets. (b-d) The sliding assembly of two graphene nanosheets (b) at initial orientation with separated distance of 0.7 nm in *z*-direction, (c) with two half-assembled snapshots (T=80 ps, 160 ps), and (d) with the formed graphene-water-graphene sandwiched structure. The displayed water within the overlapped region represents the interfacial water within the distance of 0.5 nm to both graphene nanosheets.

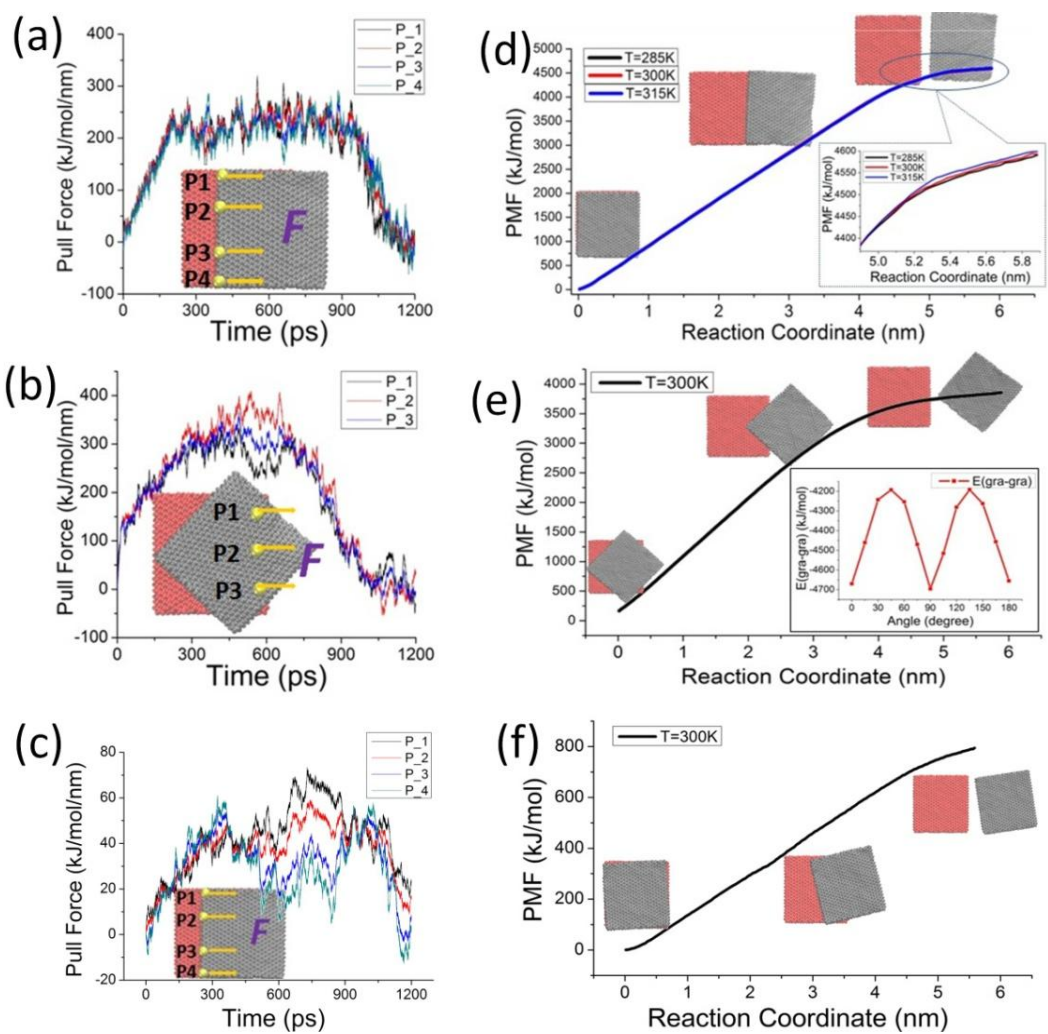

Figure 4. (a-c) The force spectra for sliding assembly with (a) edge-edge contact ($\theta = 0°$), (b) edge-corner contact ($\theta = 45°$) and (c) indirect contact ($d_z$=0.7 nm) assembly pathways. Pull groups and the applied pull forces were illustrated as insets. (d-f) PMFs and intermediate states of two graphene nanosheets in (d) edge-edge contact, (e) edge-corner contact and (f) indirect contact sliding assembly. The inset panel in (e) represents the interaction energies between two graphene nanosheets.

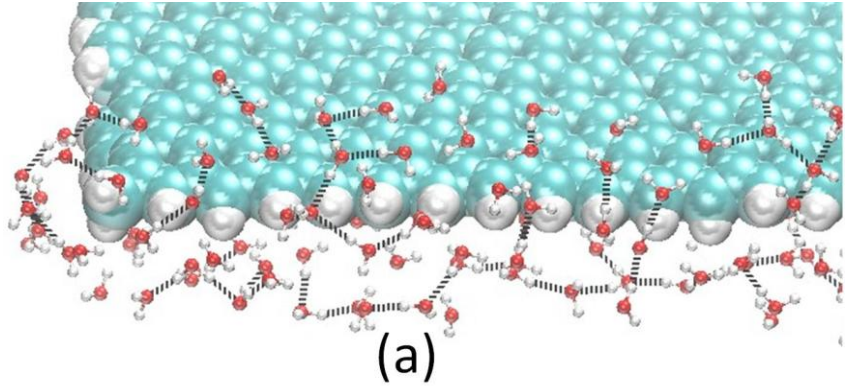

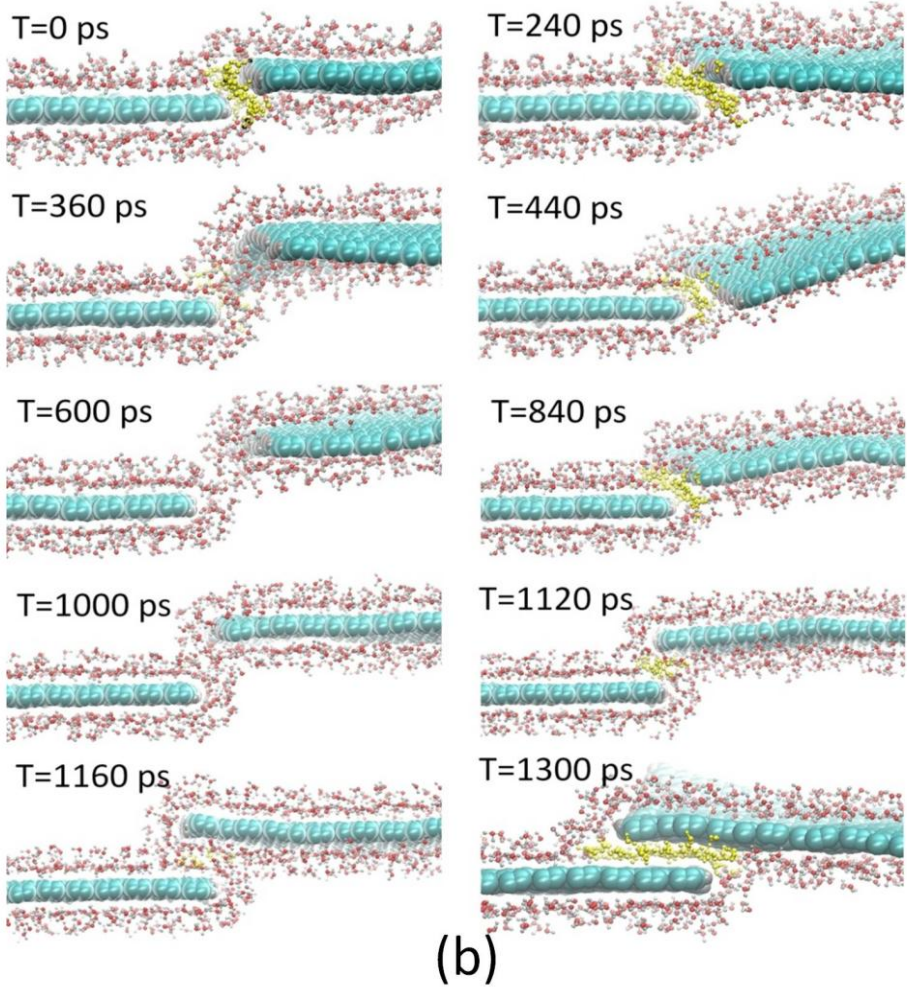

Figure 5. (a) A snapshot of hydration shell at graphene edge with H-bonds drawn as black dot lines. (b) The evolution of hydration shell during the sliding assembly of two graphene nanosheets (Sim2). The interfacial water within the distance of 0.6 nm to graphenesheet) are drawn in "CPK model"; the hindering water in contact region within the distance of 0.5 nm to both graphene nanosheets) are highlighted in yellow; graphene nanosheets are drawn in "vdW model", colored cyan.

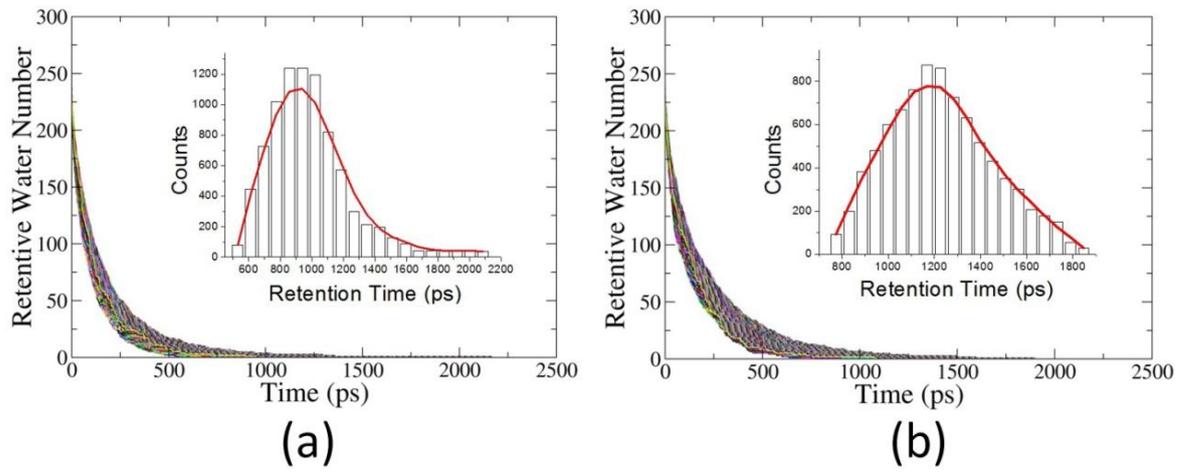

Figure 6. The evolution of retentive water number for MWF in (a) 1D and (b) 3D restraint with statistical distribution of retention time.

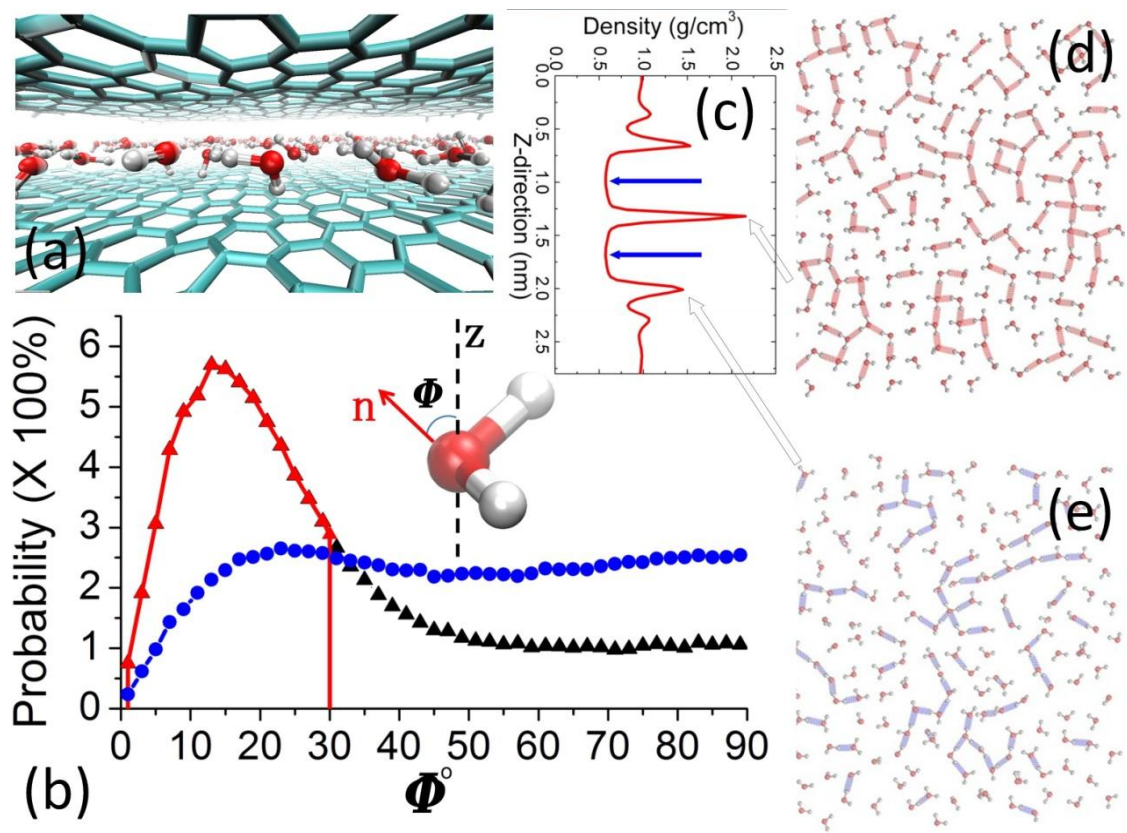

Figure 7. The structural and H-bonds properties of MFW. (a) A snapshot of MWF within interplate of two graphene nanosheets; (b) The distribution of molecular orientation for water in MWF (triangle) and the unconfined interfacial water (dot); (c) The density profile of water in $z$-direction. The blue arrow notes the position of graphene. In the right panel, the status of H-bonds of two high density water layers, (d) MWF and (e) interfacial water (first layer of water at the out surface of graphene) are showed.

Table 1. The average number of H-bonds per water ($<n>$) in MWF, Interfacial water and Block water. The internal H-bonds and the external H-bonds are the average numbers of H-bonds within the analyzed water layers, and between the analyzed water layer and the ambient water, respectively.

|  | Internal H-bonds | External H-bonds | Water molecules | $<n>$ |
|---|---|---|---|---|
| **MWF** | 243.86±10.06 | 50.60±5.22 | 189.95±3.21 | 2.83±0.08 |
| **Interfacial** | 207.33±10.65 | 161.36±8.69 | 191.23±4.97 | 3.01±0.08 |
| **Block** | 233.10±12.62 | 302.66±12.06 | 234.07±6.63 | 3.28±0.07 |